\documentclass[sigconf]{acmart}
\acmConference[ICSE 2022]{The 44th International Conference on Software Engineering}{May 21–29, 2022}{Pittsburgh, PA, USA}
\usepackage{multirow}
\usepackage{setspace}
\usepackage{graphicx}
\usepackage{color,framed}
\usepackage{xcolor}
\usepackage{mathrsfs}
\usepackage{tabularx}
\usepackage{courier}
\usepackage{soul}
\usepackage{subcaption}
\usepackage{amsthm,amsmath}
\usepackage{balance}
\usepackage{hyperref}
\usepackage{float}
\usepackage{url}
\usepackage{enumitem}
\usepackage{booktabs}
\usepackage{threeparttable}
\usepackage{comment}

\usepackage[most]{tcolorbox}
\definecolor{cGreen}{RGB}{160,238,225}
\definecolor{cRed}{RGB}{248,149,136}
\definecolor{cYellow}{RGB}{249,226,100}
\definecolor{cPatch}{RGB}{135,232,133}

\AtBeginDocument{%
  \providecommand\BibTeX{{%
    \normalfont B\kern-0.5em{\scshape i\kern-0.25em b}\kern-0.8em\TeX}}}

\setcopyright{acmcopyright}
\copyrightyear{2022}
\acmYear{2022}
\setcopyright{acmcopyright}\acmConference[ICSE '22]{44th International Conference
on Software Engineering}{May 21--29, 2022}{Pittsburgh, PA, USA}
\acmBooktitle{44th International Conference on Software Engineering (ICSE '22), May
21--29, 2022, Pittsburgh, PA, USA}
\acmDOI{10.1145/3510003.3510219}

\acmPrice{15.00}
\acmISBN{978-1-4503-9221-1/22/05}

\begin{document}
\newcolumntype{C}[1]{>{\centering\arraybackslash}p{#1}}
\title{\emph{MVD}: Memory-Related Vulnerability Detection Based on Flow-Sensitive Graph Neural Networks}
\thanks{\\
*Xiaobing Sun and Lili Bo are the corresponding authors.\\
}
\author{Sicong Cao}
\affiliation{%
  \institution{Yangzhou University}
  \city{Yangzhou}
  \country{China}
}
\email{MX120190439@yzu.edu.cn}

\author{Xiaobing Sun}
\authornotemark[1]
\affiliation{%
  \institution{Yangzhou University}
  \city{Yangzhou}
  \country{China}}
\email{xbsun@yzu.edu.cn}

\author{Lili Bo}
\authornotemark[1]
\affiliation{%
  \institution{Yangzhou University}
  \city{Yangzhou}
  \country{China}}
\email{lilibo@yzu.edu.cn}
  
\author{Rongxin Wu}
\affiliation{%
  \institution{Xiamen University}
  \city{Xiamen}
  \country{China}}
\email{wurongxin@xmu.edu.cn}

\author{Bin Li}
\affiliation{%
  \institution{Yangzhou University}
  \city{Yangzhou}
  \country{China}}
\email{lb@yzu.edu.cn}

\author{Chuanqi Tao}
\affiliation{%
  \institution{Nanjing University of Aeronautics and Astronautics}
  \city{Nanjing}
  \country{China}}
\email{taochuanqi@nuaa.edu.cn}

\begin{abstract}
\emph{Memory-related vulnerabilities} constitute severe threats to the security of modern software. 
Despite the success of deep learning-based approaches to generic vulnerability detection, 
they are still limited by the underutilization of flow information  
when applied for detecting memory-related vulnerabilities, leading to high false positives. 

In this paper, we propose \emph{MVD}, a \emph{statement-level} \underline{\textbf{M}}emory-related \underline{\textbf{V}}ulnerability \underline{\textbf{D}}etection approach based on flow-sensitive graph neural networks (FS-GNN).  FS-GNN is employed to jointly embed both unstructured information (i.e., source code) and structured information (i.e., control- and data-flow) to capture implicit memory-related vulnerability patterns. We evaluate \emph{MVD} on the dataset which contains 4,353 real-world memory-related vulnerabilities, and compare our approach with three state-of-the-art deep learning-based approaches as well as five popular static analysis-based memory detectors.  The experiment results show that \emph{MVD} achieves better detection accuracy, outperforming both state-of-the-art DL-based and static analysis-based approaches. Furthermore, \emph{MVD} makes a great trade-off between accuracy and efficiency.

\end{abstract}


\maketitle

\section{Introduction}
As one of the most representative vulnerabilities, \emph{memory-related vulnerabilities} can result in performance degradation and program crash, severely threatening the security of modern software \cite{DBLP:journals/smr/WeiSBCXL21,SMATUS}. According to the data released by CVE (Common Vulnerabilities and Exposures \cite{CVE}), nearly a third of the vulnerabilities (32.6\%) in Linux Kernel \cite{Linux} are related to improper memory operations \cite{DBLP:journals/corr/abs-2104-04385}. 

Many static analysis approaches \cite{SMOKE,PCA,Saber1,Saber2,FastCheck,Pinpoint,DBLP:conf/internetware/JiYXFL12,DBLP:conf/sas/OrlovichR06,DBLP:conf/icse/HeineL06,CBMC,DBLP:journals/tse/0001JMCL19} have been proposed to detect memory-related vulnerabilities and shown their effectiveness. 
They 
use some pre-defined vulnerability rules or patterns to search for improper memory operations \cite{Pointer1,Pointer2}. 
However, well-defined vulnerability rules or patterns are highly dependent on expert knowledge,
and thus it is difficult to cover all the cases. 
What's worse, the sophisticated programming logic in real-world software projects gets in the way of the manual identification of the rules, and thus greatly compromises the performance of the traditional static analysis-based approaches \cite{Empirical2}. 
Recently, benefiting from the powerful performance of deep learning (DL), a number of approaches \cite{FUNDED,VulDeePecker,DBLP:journals/infsof/ZhouSXLC19,mVulDeePecker,Devign,SySeVR,DeepWukong,TOKEN,REVEAL,IVDETECT,BGNN4VD} have been proposed to leverage DL models to capture program semantics to identify potential software vulnerabilities. Compared with traditional static analysis-based approaches, they can automatically extract implicit vulnerability patterns from prior vulnerable code instead of requiring expert involvement.
However, the existing DL-based approaches suffer from two limitations when applied to memory-related vulnerability detection, as described below. 

\textbf{\emph{Flow Information Underutilization:}} 
Due to the underutilization of flow information, existing DL-based approaches failed to detect   complicated memory-related vulnerabilities in real-world projects \cite{DeepWukong} for the following two aspects: (1) lack of interprocedural analysis, and (2) partial flow information loss in model training. For the former one, most of DL-based approaches \cite{Devign,TOKEN,REVEAL,IVDETECT,FUNDED} take the \emph{function-level} vulnerable code as input to conduct intraprocedural analysis for feature extraction, ignoring call relations between functions. However, in real-world programs, operations like calling a user-defined function which realizes memory allocation or free are widespread. Missing interprocedural analysis may cause incomplete semantic modeling, resulting in lower Recall and Precision. For the latter one, limited by the capability of popular DL models (e.g., BiLSTM \cite{VulDeePecker,DBLP:journals/chinaf/SunPZLC19,SySeVR,DBLP:conf/iceccs/GongXLFH19}, GGNN \cite{FUNDED,Devign}, and GCN \cite{DeepWukong}) in handling multiple relations, partial flow information is lost during the process of model training. For example, \emph{Devign} \cite{Devign} uses GGNN \cite{GGNN} as the basic model to propagate and aggregate information across multi-relational graphs. Since GGNN treats the relational graph as multiple directed graphs without attributes (i.e., feature information is only passed between nodes connected by edges of the same type), its effectiveness is often compromised by the tremendous increase in the number of data-flow edges with different attributes. Thus, \emph{Devign} has to substitute them with three other \emph{token-level} relations (i.e., \emph{LastRead}, \emph{LastWrite}, and \emph{ComputedFrom} \cite{DBLP:conf/iclr/AllamanisBK18}) to make it more adaptive for the graph embedding, sacrificing partial precise data-flow information well preserved in graphs. A simple instance is that receiving a normal pointer variable (non-vulnerable) is obviously not the same as receiving a pointer variable which points to the memory just released (vulnerable).

\textbf{\emph{Coarse Granularity:}} The detection granularity of the existing DL-based approaches is mostly at the \emph{function-level} \cite{Devign,FUNDED,TOKEN,REVEAL,IVDETECT} or \emph{slice-level} \cite{VulDeePecker,SySeVR,DeepWukong,mVulDeePecker}.
However,  developers still need to spend a  deal of time in manually narrowing down the range of suspicious statements (or operations). 
Achieving fine-grained detection results is non-trivial. 
Due to the huge differences between various vulnerabilities, existing DL-based approaches for generic vulnerability detection have to sacrifice unique semantic features specific to certain vulnerabilities to ensure that the trained model can cover the general characteristics of the  majority of vulnerabilities. 
In comparison with other vulnerabilities,  memory-related vulnerabilities  are usually fixed with one or several lines of code, which makes fine-grained detection possible. For example, \emph{memory leak} can be located to the statement which allocates memory, while \emph{use-after-free} can be located to the statement which frees memory.

In this paper, we propose  a novel approach (\emph{MVD}) based on flow-sensitive graph neural networks to alleviate the above limitations.

\textbf{\emph{Fully Utilizing Flow Information:}} To capture more comprehensive and precise program semantics, \emph{MVD} combines Program Dependence Graph (PDG) with Call Graph (CG) \cite{CG} to capture interprocedural control- and data-flow information. First, we conduct interprocedural analysis by extending PDG with additional semantic information (including call relations and return values between functions) 
using CG. In our approach, code snippets and relations (i.e., edges) are embedded in compact low-dimensional representations to preserve both the unstructured (i.e., source code) and structured (i.e., control- and data-flow) information. 
Furthermore, in order to make the detection model learn effective memory-related vulnerability patterns from comprehensive and precise flow information, \emph{MVD} constructs a novel Flow-Sensitive Graph Neural Networks (FS-GNN) to jointly embed statements and flow information to capture program semantics from vulnerable code.

\textbf{\emph{Fine Granularity:}} 
We formalize the detection of vulnerable statements as a node classification problem, i.e., identifying which statement(s) in the program is vulnerable. Specifically, \emph{MVD} receives the graph representation  of a program (in which graph nodes represent statements and edges indicate their  relations) and outputs node labels (i.e., vulnerable or not). 

Since there is currently no dataset that can be directly used for training a \emph{statement-level} memory-related vulnerability detection model, we   construct a dataset which contains 4,353 real-world memory-related vulnerabilities. The dataset as well as the empirical data are available online\footnote{\href{https://github.com/MVDetection/MVD}{https://github.com/MVDetection/MVD}}.

In summary, this paper makes the following contributions:
\begin{itemize}[leftmargin=2em]
    \item We propose a novel Flow-Sensitive Graph Neural Networks (FS-GNN) to support effective detection of   memory-related vulnerabilities.
    \item We formalize vulnerability detection as a fine-grained node classification problem to identify suspicious vulnerable statements.
    \item We evaluate \emph{MVD} on our constructed dataset, and the results show that \emph{MVD} can effectively detect memory-related vulnerabilities over state-of-the-art  vulnerability detection approaches (including three DL-based and five static analysis-based approaches).
\end{itemize}


\section{Basics and Motivation}
\subsection{Definitions}

\textbf{Program Dependence Graph.} Given a program, all the program statements and dependencies among statements constitute a \emph{Program Dependence Graph (PDG)} \cite{PDG}. \emph{PDG} includes two types of edges: data
dependency edges which reflect the influence of one variable on another and control dependency edges which reflect the influence of predicates on the values of variables.

\textbf{Call Graph.} Given a program, its \emph{Call Graph (CG)} \cite{CG} indicates a series of function calls from call sites (caller) to the callee. 

\textbf{Graph Neural Networks.} Due to the outstanding ability in processing graph data structures, \emph{Graph Neural Networks (GNNs)} have been used in a variety of data-driven software engineering (SE) tasks (e.g., code representation \cite{DBLP:conf/iclr/AllamanisBK18}, clone detection \cite{DBLP:conf/wcre/WangLM0J20}, and bug localization \cite{DBLP:conf/sigsoft/LouZDLSHZZ21}) and have achieved great breakthroughs. The goal of GNNs is to train a parametric function via message passing between the nodes of graphs for downstream tasks, i.e., graph classification, node classification, and link prediction.

\subsection{Motivating Examples}\label{Motivation}

\begin{figure}
  \centering
  \includegraphics[width=\linewidth]{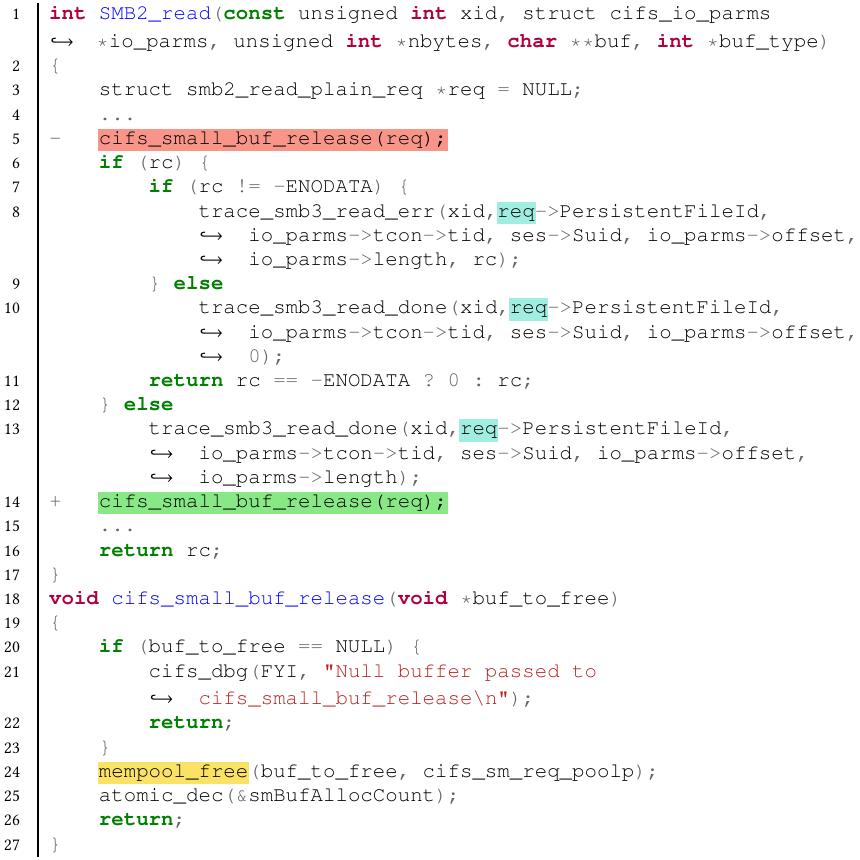}
\caption{A \emph{Use-After-Free} Vulnerability (CVE-2019-15920) in \emph{Linux Kernel}}
\label{1}
\end{figure}

\begin{figure*}[h]
  \centering
  \includegraphics[width=\linewidth]{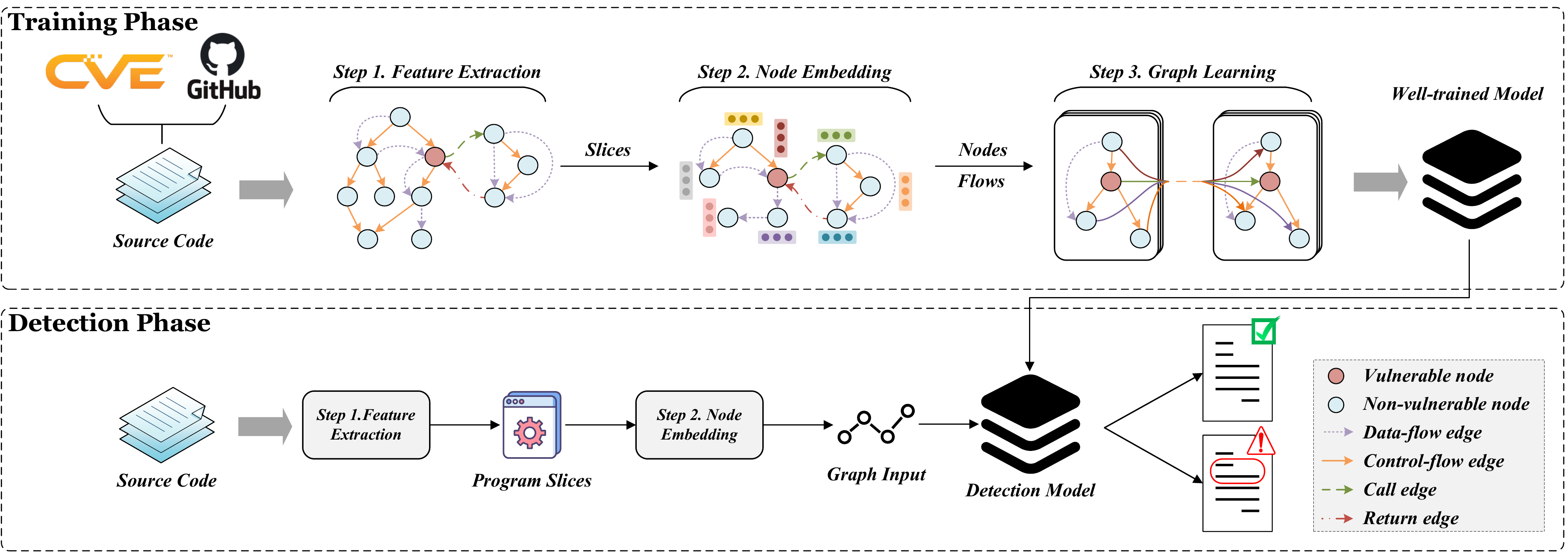}
\caption{Overview of \emph{MVD}}
\label{2}
\end{figure*}

Figure \ref{1} shows a typical \emph{use-after-free} vulnerability CVE-2019-15920 \cite{CVE-2019-15920} in \emph{Linux Kernel}. The vulnerable function \texttt{SMB2\_read} has been simplified for a clear illustration. We can observe that the memory space pointed by the pointer \texttt{req} is released in advance by the memory release statement \texttt{cifs\_small\_buf\_release(req)} at \underline{line 5}, while it is still used at \underline{lines 8-13}. This operation may allow attackers to write malicious data. To fix this vulnerability, the pointer \texttt{req} should be released after it is used for the last time (e.g., \underline{line 14}). Despite the support of precise data dependence analysis, this vulnerability cannot be easily detected by some static analysis-based approaches because they may not know \texttt{mempool\_free()} at \underline{line 24} is a user-defined memory deallocation function.

From this example, we can draw the following observations:

\textbf{Observation 1. Comprehensive and precise interprocedural flow analysis is necessary.} As shown in Figure \ref{1}, we can find that the program semantics of vulnerable code and non-vulnerable code are different. In the vulnerable code, vulnerable statement at \underline{line 5} releases \texttt{req} when \texttt{req} is still being used after that, while in the non-vulnerable code, \texttt{req} is released by the patched statement at \underline{line 14} only when \texttt{req} is no longer  used. However, due to the lack of interprocedural analysis, critical program semantics (i.e., memory deallocation via \texttt{mempool\_free}, which is involved in the function call to \texttt{cifs\_small\_buf\_release(req)} at \underline{line 5}) are ignored by a number of deep learning-based approaches \cite{REVEAL,Devign,FUNDED,IVDETECT}, resulting in incomplete program semantic modeling towards vulnerable statement at \underline{line 5}. 

In our approach, we extend basic Program Dependence Graph (PDG) with additional semantic information like call relations and return values obtained from Call Graph (CG) \cite{CG} to capture comprehensive and precise interprocedural program semantics. 
With such rich information, features of memory-related vulnerabilities can be  extracted for more effective detection.

\textbf{Observation 2. Sensitive contextual information within flows helps to refine detection granularity.} As shown in the motivating example, the premise of identifying the statement at \underline{line 5} as vulnerable is that we should know in advance that \texttt{req} released by this statement will be used later. Thus, to distinguish vulnerable statements from others, the neural networks used as detection models should be able to capture sensitive contextual information within flows of vulnerable statements for inference. However, due to the limitations of popular DL models\cite{SySeVR,VulDeePecker,FUNDED,Devign} in handling multiple relations, rich contextual information are lost during the process of model training. For example, \emph{FUNDED} \cite{FUNDED} only considers one-directional transmission of multiple relations such as control- and data-flows, and adopts GGNN \cite{GGNN} to learn vulnerability patterns for detection. While it is successful in \emph{function-level} vulnerability detection, it loses some important contextual information from output flows, e.g., the data-flow information within Edge 5->8 will not be used for feature update of vulnerable statement at \underline{line 5}. Thus, it is hard for \emph{FUNDED} to identify the statement \underline{line 5} as vulnerable because critical output flow information (i.e., using \texttt{req} after freeing it) is not preserved in \texttt{cifs\_small\_buf\_release(req)}.

Based on the above observations,  we propose a novel model, FS-GNN, for effectively detecting memory-related vulnerabilities. FS-GNN is a novel flow-sensitive graph neural network to jointly embed both statements and flow information for better information propagation between statements. With FS-GNN, rich contextual semantics of neighbors are aggregated through multiple relations to update the embedding of the central node.

\section{Our Approach: \emph{MVD}}\label{Approach}
Figure \ref{2} shows the overview of \emph{MVD}. It consists of two phases:  training phase and  detection phase. 

The training phase includes three steps. In step 1, \emph{MVD} constructs the Program Dependence Graph (PDG) based on the control- and data-flow of the program. To capture comprehensive and precise program semantics, \emph{MVD} extends the PDG with additional semantic information like call relations and return values obtained from Call Graph (CG) \cite{CG} to conduct interprocedural analysis. Furthermore, to reduce irrelevant semantic noise, \emph{MVD} conduct program slicing \cite{PS,DBLP:journals/ese/SoremekunKBZ21} from the program points of interest. In step 2, \emph{MVD} uses Doc2Vec \cite{Doc2Vec} to transform the statements of each slice into low-dimensional vector representations. In step 3, \emph{MVD} uses Flow-Sensitive Graph Neural Networks (FS-GNN) to jointly embed nodes and relations to learn implicit vulnerability patterns and re-balance node labels distribution. Finally, a well-trained model is produced for memory-related vulnerability detection at the \emph{statement-level}.

For the detection phase, with the interprocedural analysis, the control- and data-dependence of a target program is first extracted for program slicing to capture precise program semantics related to memory usage. Then, for each slice, both its unstructured (i.e., statement embedding by Doc2Vec) and structured (i.e., control- and data-flow) information are used as graph input to feed into the well-trained detection model for vulnerability detection.

\begin{figure}
\centering
\begin{subfigure}{\linewidth}
\centering
\includegraphics[width=\linewidth]{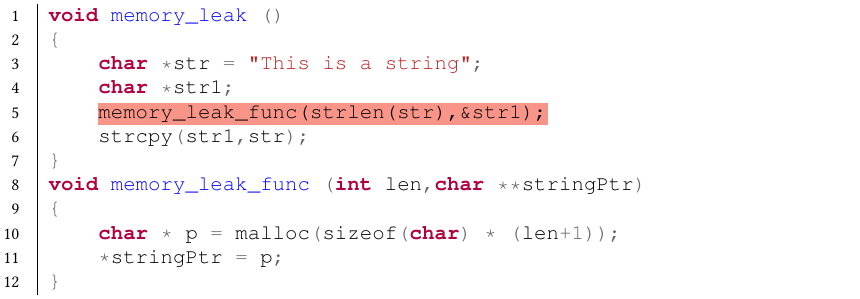}
\caption{Exemplary Code Sample}\label{3a}
\quad
\end{subfigure}      
\begin{subfigure}{\linewidth}
\centering   
\includegraphics[width=\linewidth]{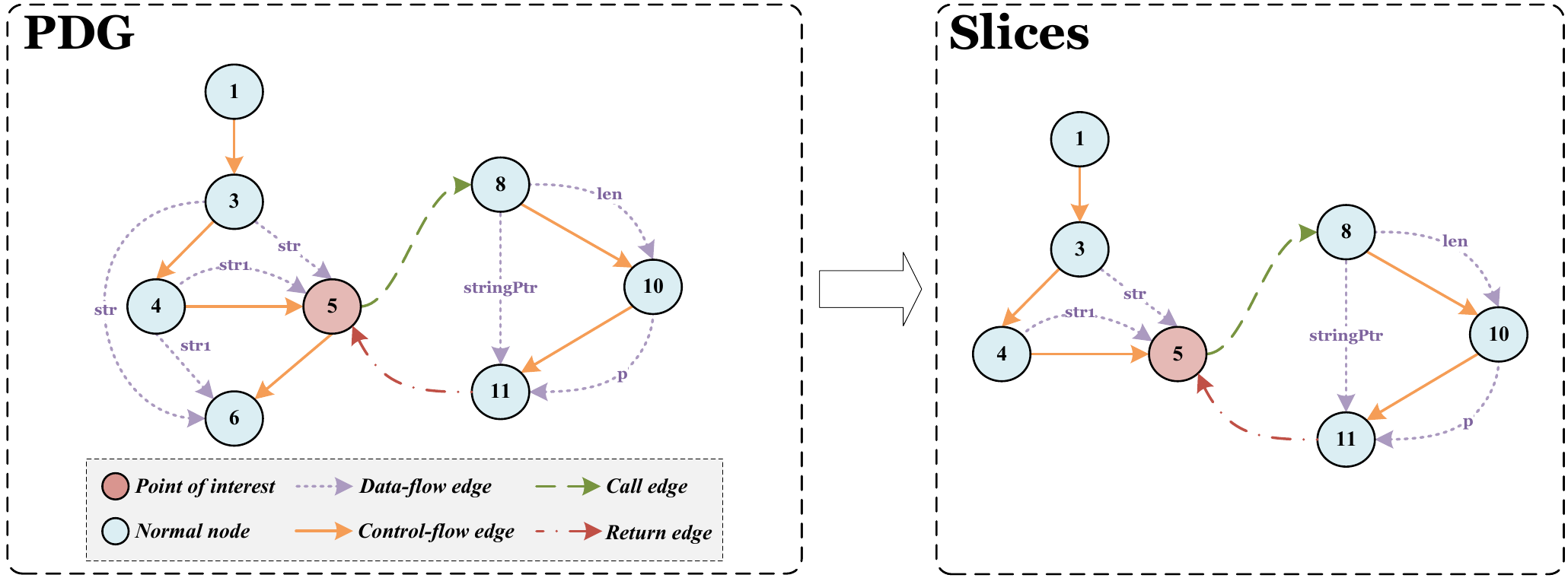}
\caption{Program Slicing}\label{3b}
\end{subfigure}
\caption{
\label{3}Details of Feature Extraction}
\end{figure}

\subsection{Feature Extraction}\label{Feature Extraction}
First, we use the static analysis tool, \emph{Joern} \cite{DBLP:conf/sp/YamaguchiGAR14}, to parse source code and construct the Program Dependence Graph (PDG). Then, we extend the PDG with additional semantic information like call relations and return values obtained from Call Graph (CG) to conduct interprocedural analysis, which preserves comprehensive control- and data-flow information.  However, since a function usually contains dozens or even hundreds of code lines while the vulnerability exists only in a few lines of code, simply taking the whole program to train a detection model will reduce the performance of identifying key features. Thus, \emph{MVD} adopts program slicing \cite{PS} to perform backward and forward slicing from the program point of interest to avoid noise induced by irrelevant statements. 

To ensure that the  generated slices   contain memory-related vulnerabilities, we mainly focus on two types of program points of interest: 1) \emph{system API calls}, and 2) \emph{pointer variable}. As mentioned in previous works \cite{VulDeePecker,SySeVR,mVulDeePecker,DeepWukong}, the misuse of \emph{system API calls} is one of the major causes of vulnerabilities, including memory-related vulnerabilities. For example, \emph{syscall\_buf} is a typical \emph{system API call} related to buffer operations in \emph{Linux Kernel}. It often occurs in \emph{Out-of-bounds Read/Write} and other similar buffer-related vulnerabilities. In total, we conclude 217 \emph{system API calls} from several static memory detectors \cite{Pinpoint,SMOKE} as slicing criteria for extracting vulnerable code snippets. For \emph{pointer variable}, it has been widely adopted by traditional static analysis-based approaches \cite{Saber1,PCA,SMOKE}. It should be noted that starting from the program point of interest, we perform backward slicing according to both control- and data-dependence, but forward slicing based on only data-dependence because improper memory operations (e.g., allocating memory but not freeing it) have been involved in the forward data-dependence, and usually forward control-dependence will induce a great deal of irrelevant statements. 

Figure \ref{3} provides  an example to show the process of  feature extraction. As shown in Figure \ref{3a}, it is a \emph{memory leak} vulnerability. At \underline{line 5}, it allocates memory through \texttt{malloc()} in function \texttt{memory\_leak\_func()} without freeing even to the end of the program. In our approach, the control- and data-flow information of the vulnerable program is first extracted to construct the PDG of the program, which is shown in Figure \ref{3b}. Then, based on interprocedural analysis, the call relation (i.e., Edge 5->8) and return value (i.e., Edge 11->5) information is added. To reduce irrelevant nodes, we adopt the sensitive function call at \underline{line 5} (i.e., Node 5 highlighted in red) as the program point to perform backward and forward slicing.  Node 6 is control dependent on Node 5 with an Edge 5->6, and data dependent on Node 4 with an Edge 4->6. After slicing, Node 6 is removed because it is not data-dependent on Node 5.

\subsection{Node Embedding}\label{Node Embedding}
After feature extraction, we transform all statement nodes in the graph into low-dimensional vectors as input to the graph neural network models.

We use Doc2Vec \cite{Doc2Vec} to represent code statements as vectors since it is a widely-used technique \cite{DeepWukong,DBLP:journals/isci/GhaffarianS21} to encode the documents (i.e., code statements) instead of an individual word (e.g., a variable) into a fixed-length vector. 
Then, the vectors of the statement and context words are input to the hidden layer to obtain the intermediate vector as the input of \emph{softmax} layer. Finally, in the inference stage, the vector of a given statement is achieved through Stochastic Gradient Descent (SGD) \cite{SGD}. In this way, the Doc2Vec can provide a more precise embedding of the code statement that will preserve the semantic information.

\subsection{Graph Learning}\label{Graph Learning}
\begin{figure*}
  \centering
  \includegraphics[width=\linewidth]{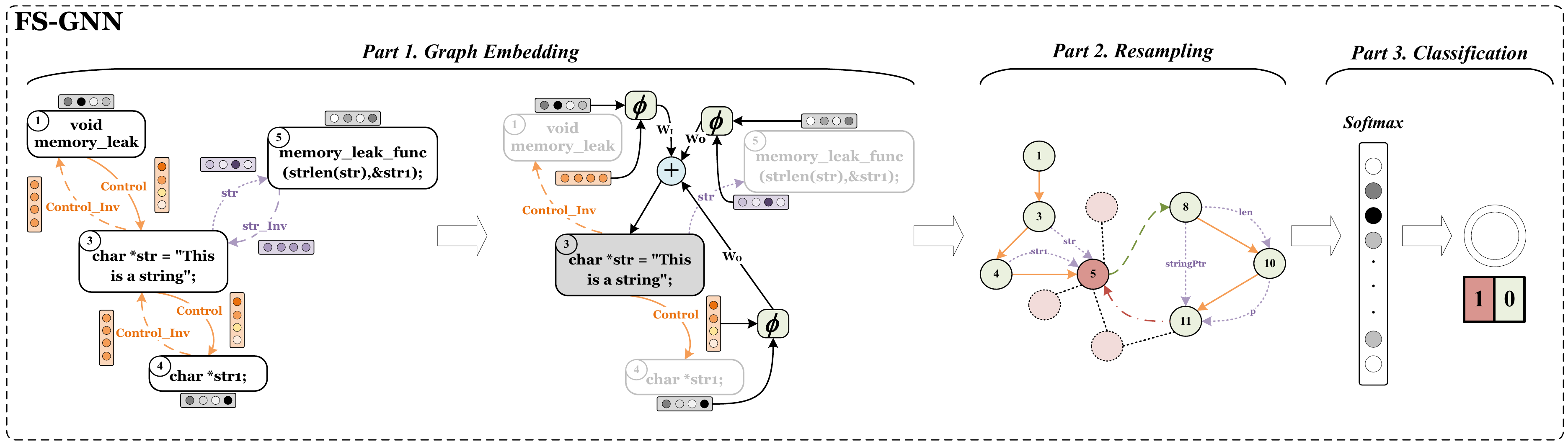}
\caption{Graph Learning with FS-GNN}
\label{4}
\end{figure*}

To train a model which can learn implicit vulnerability patterns from source code and identify suspicious vulnerable statements, we construct a novel graph learning framework, Flow-Sensitive Graph Neural Network (FS-GNN) for graph learning. The details of our approach are shown in Figure \ref{4}. The key insight of FS-GNN is to jointly embed both statement embedding and flows information to capture sensitive contextual information for  semantic learning. FS-GNN is composed of three parts: graph embedding, resampling and  classification.

\textbf{Graph Embedding.}
Different from most of the existing graph embedding approaches that embed only nodes in the graph,  we leverage the entity-relation composition operations $\phi(\cdot)$ used in Knowledge Graph embedding approaches \cite{DBLP:conf/nips/BordesUGWY13} to jointly embed statement nodes and multiple flow edges to incorporate edge embedding into the update of node information.
To be specific, during the process of graph embedding in FS-GNN, the node embedding $h_v$ of statement node $v$ can be updated by:
\begin{equation}
\begin{split}
    \boldsymbol{h}_v = f\left({\sum\limits_{(u,r)\in \mathcal{N}(v)}\boldsymbol{W}_{\lambda(r)}\phi(\boldsymbol{x}_u,\boldsymbol{z}_r)}\right)
\end{split}
\end{equation} 

where $\boldsymbol{h}_v$ denotes the updated representation of node $v$. $\mathcal{N}(v)$ is a set of immediate neighbors of \emph{v} for its outgoing edges. $\phi(\cdot)$ is a composition operator, including subtraction, multiplication, and circular-correlation. $\boldsymbol{x}_u$ and $\boldsymbol{z}_r$ denotes initial features for node $u$ (encoded by Doc2Vec) and edge $r$, respectively. Similar to traditional Relational Graph Neural Networks (RGCN) \cite{RGCN}, initial edge representation for edge $r$ can be encoded by basis decomposition \cite{RGCN} as $\boldsymbol{z}_r = \sum\limits_{b=1}^{\mathcal{B}}\alpha_{br}\boldsymbol{v}_b$, where $\boldsymbol{v}_{b\in\mathcal{B}}$ is a set of learnable basis vectors and $\alpha_{br}\in\mathbb{R}$ is also the learnable scalar weight specific to edge type and basis. $\boldsymbol{W}_{\lambda(r)}$ represents a edge type specific parameter. To make FS-GNN context-aware and capture important information from outgoing edges, we double edges by adding inverse edges and assign different weight parameters according to edge types (i.e., $\boldsymbol{W}_{\lambda(r)} = \boldsymbol{W}_O$ when $r$ is an initial edge, and $\boldsymbol{W}_{\lambda(r)} = \boldsymbol{W}_I$ when $r$ is an inverse edge).

Similarly, the edge embedding $h_r$ of edge $r$ can be updated by $\boldsymbol{h}_r =\boldsymbol{W}_{rel}\boldsymbol{z}_r$, where $\boldsymbol{W}_{rel}$ is a learnable transformation matrix which projects all the relations to the same embedding space as nodes.

Finally, the representation of a node $v$ and edge $r$ updated after $l$ layers are shown as:
\begin{equation}\label{form4}
\begin{split}
    \boldsymbol{h}_v^{l+1} &= f\left({\sum\limits_{(u,r)\in \mathcal{N}(v)}\boldsymbol{W}_{\lambda(r)}^l\phi(\boldsymbol{h}_u^l,\boldsymbol{h}_r^l)}\right)
\end{split}
\end{equation}

\begin{equation}
\begin{split}
    \boldsymbol{h}_r^{l+1} &=\boldsymbol{W}_{rel}^l\boldsymbol{z}_r^l
\end{split}
\end{equation}

Note that $h_v^0=x_u$ and $h_r^0=z_r$ (i.e., initial representation of node $v$ and edge $r$).

With the help of our flow-sensitive graph learning, contextual information can be captured  and sensitive flow information is given more attention. For example, in Figure \ref{4}, initial node representation is encoded by Doc2Vec and edge representation is calculated by basis decomposition. Edge matrix is inversed first to capture contextual feature information. Then, to aggregate information from neighbor nodes to update the representation of Node 3, initial representations of Nodes 1, 4, 5 are embedded jointly with their incoming edges (i.e., Edges 3->1, 3->4, 3->5) by Eq. \ref{form4} to preserve some important features from outgoing nodes.

\textbf{Resampling.} After $l$ layers graph learning, directly training the classifiers on all statement nodes is biased because the distribution of non-vulnerable nodes and vulnerable nodes is extremely imbalanced. For example, in Figure \ref{3}, although we have filtered out some irrelevant nodes by program slicing, the number of non-vulnerable nodes (i.e., Node 1-4, 6, 8-11) is still larger than that of vulnerable nodes (i.e., Node 5). To generate some synthetic vulnerable nodes to re-balance the distribution, we adopt \emph{GraphSMOTE} \cite{GraphSMOTE}, a \emph{graph-level} oversampling framework, as the basic component for our resampling. 

Concretely, it contains two steps: (1) node generation, and (2) edge generation. Firstly, to generate high-quality synthetic nodes, we utilize the widely-used \emph{SMOTE} \cite{SMOTE} algorithm to perform interpolation on vulnerable nodes. It searches for the closest neighbour node around each minority node (i.e., vulnerable node) in the embedding space and generates synthetic nodes between them. Then, edge generator adopts weighted inner production \cite{GraphSMOTE} to generate edges and gives link predictions for synthetic nodes by setting a threshold $\eta$ to keep the connectivity of the graph. If the predicted probability of connection between synthetic node $v'$ and its closest neighbor node $u$ is greater than $\eta$, both the synthetic node $v'$ and edge $[v',u]$ will be put into the augmented adjacency matrix of original graphs. To make the analysis easier, the type of all synthetic edges is set as \emph{"Control"} (i.e., synthetic nodes are control-dependent on their neighbor nodes)\footnote{We omit data-dependency flow because during the empirical study, we find that a large number of irrelevant synthetic data-dependency edges can introduce biases and make the performance of the detection model deteriorate.}.

Owing to the contribution of resampling, the proportion of memory-related vulnerable statements increases, avoiding the well-trained detection model biased caused by imbalanced distribution of vulnerable nodes and non vulnerable nodes. For example, in Figure \ref{4}, three synthetic nodes (Pink-shaded) are connected with one vulnerable node (i.e., Node 5) and one non-vulnerable node (i.e., Node 11).

\textbf{Classification.}
Before training the classification model, FS-GNN  adopts one-layer \emph{flow-sensitive graph learning} block in Section \ref{Feature Extraction} again to update node information by Eq. \ref{form4}. By learning both the unstructured (i.e., statement embedding) and structured (i.e., various flows) features from nodes and edges, the classification model are employed to distinguish vulnerable and non-vulnerable statements.

To train the model, we use the \emph{softmax} activation function as the last linear layer for node classification and minimize the following cross-entropy loss on all labeled nodes (i.e., vulnerable or non-vulnerable):
\begin{equation}
\begin{split}
    \mathop{min}\limits_{\theta}\mathcal{L} &= -\sum\limits_{G\in \mathcal{G}}\frac{1}{\mathcal{|V|}}\sum\limits_{i\in \mathcal{V}}\sum\limits_{k=1}^K{t_{ik}}\ln{h_{ik}^{(L)}}
\end{split}
\end{equation}

where $G$ is a code slice graph in the training set $\mathcal{G}$, $\mathcal{V}$ is the set of nodes in our training set. $h_{ik}^{(L)}$ represents the probability of node $i$ belonging to class $k$, where $k=\{0,1\}$ for the binary node classification task. $t_{ik}$ denotes respective ground truth label for node $i$.

\subsection{Vulnerability Detection}
In the detection phase, we apply the well-trained model to detect potential memory-related vulnerabilities in programs and identify suspicious statements.

Specifically, similar to training phase, program semantics reflected in the graph representations of source code are captured through interprocedural analysis. In order to reduce the number of memory operations-irrelevant statements, programs are sliced according to points of interest (\emph{system API calls} and \emph{pointer variable}) to obtain a batch of program slices (Section \ref{Feature Extraction}). Next, statement nodes in program slices are embedded into low-dimensional vectors through Doc2Vec (Section \ref{Node Embedding}). Finally, both unstructured (i.e., statement embedding) and structured (i.e., control- and data-flow) information are used as graph input to feed into the well-trained detection model for vulnerability detection.

\section{EXPERIMENTS}\label{Experiments}
\subsection{Research Questions}

\noindent\textbf{RQ1.} How effective is \emph{MVD} in detecting memory-related vulnerabilities compared to existing deep learning-based vulnerability detection approaches?
\
\newline
\indent 
Works most relevant to \emph{MVD} are deep learning-based vulnerability detection approaches. By investigating this RQ, we aim to answer how well does \emph{MVD} perform in comparison with the state-of-the-art deep learning-based approaches in memory-related vulnerability detection.

\noindent\textbf{RQ2.} How effective is \emph{MVD} in detecting memory-related vulnerabilities compared to  static analysis-based vulnerability detectors?
\
\newline
\indent Static analysis-based vulnerability detection tools are widely-used and perform well on memory-related vulnerabilities. In addition, static analysis-based approaches can identify the  statement-level results for vulnerability detection (i.e., fine-grained detection results). Therefore, the purpose of this RQ is to analyze how \emph{MVD} perform compared with existing static analysis-based detectors.

\noindent\textbf{RQ3.} How effective is FS-GNN for memory-related vulnerability detection?
\
\newline
\indent One of the key contributions of our approach is Flow-Sensitive Graph Neural Network, which jointly embeds both unstructured (i.e., code snippets) and structured (i.e., control- and data-flows) information to learn comprehensive and precise program semantics. Different from RQ1, we aim to show whether sensitive contextual information captured by FS-GNN contributes to memory-related vulnerability detection in comparison with other popular GNNs (i.e., evaluating the effectiveness of fully utilizing flow information).

\noindent\textbf{RQ4.} How efficient are \emph{MVD} and baselines in terms of their time cost for detecting memory-related vulnerabilities?
\
\newline
\indent Efficiency is  important   for evaluating the performance of memory-related vulnerability detection approaches. An approach which costs too much time for detecting vulnerabilities may encounter adoption barriers in practice. This RQ is to investigate whether \emph{MVD} can make a better trade-off between accuracy and efficiency.

\subsection{Experiment Setup}
\subsubsection{Dataset}
Since existing vulnerability datasets are either not tailored for memory-related vulnerabilities \cite{FAN,REVEAL,DeepWukong,Devign,VulinOSS}, or not sufficient for training deep learning models (e.g., SPEC CINT2000 \cite{SPEC}), we \emph{manually} constructed a new vulnerability dataset which covers 13 common memory-related vulnerabilities (including CWE-119, -120, -121, -122, -124, -125, -126, -401, -415, -416, -476, -787, and -824) for model training and evaluation. Our dataset is based on two widely-adopted sources: (1) SARD \cite{SARD}, a well-known sample vulnerability data set, and (2) CVE \cite{CVE}, a famous vulnerability database. In this work, we focused on C/C++ programs due to their frequent memory problems caused by low-level control of memory \cite{SoK} and adopted vulnerability types (i.e., CWE-IDs \cite{CWE}) as our search criteria to collect memory-related vulnerabilities from SARD and CVE. For real-world vulnerabilities, we only considered CVEs which contain source code and from which we collect both vulnerable functions and corresponding patched functions. For SARD, we collected all test cases labeled as "bad".

\begin{table}[t]
 \caption{Details of vulnerability dataset.}
  \centering
  \begin{tabular}{ccccc}
    \toprule
    \textbf{Project}     & \textbf{\#Version} & \textbf{\#Samples}  & \textbf{\#Vertices} & \textbf{\#Edges}\\
    \midrule
    Linux Kernel &2.6-4.20  &868      &26,917 &29,512 \\
    FFmpeg       &0.5-4.1  &73        &1,971  &2,168  \\
    Asterisk     &1.4-16.14  &18        &468    &502   \\
    Libarchive   &2.2-3.4  &11          &235    &269    \\
    Libming      &0.4.7   &7           &119    &141    \\
    LibTIFF      &3.8-4.0  &24        &584    &639    \\
    Libav        &12.3  &16        &526    &573       \\
    LibPNG       &1.0.x-1.6.x &13          &392    &447 \\
    QEMU         &0.9-4.3  &121      &4,711  &5,308       \\
    Wireshark    &1.2-3.2  &57        &2,056  &2,190       \\
    SARD     &-    &3,145   &11,237     &13,049     \\
    Total        &- &4,353    &49,216 &54,798       \\
    \bottomrule
  \end{tabular}
  \label{table1}
\end{table}

\begin{figure}
  \centering
  \includegraphics[width=\linewidth]{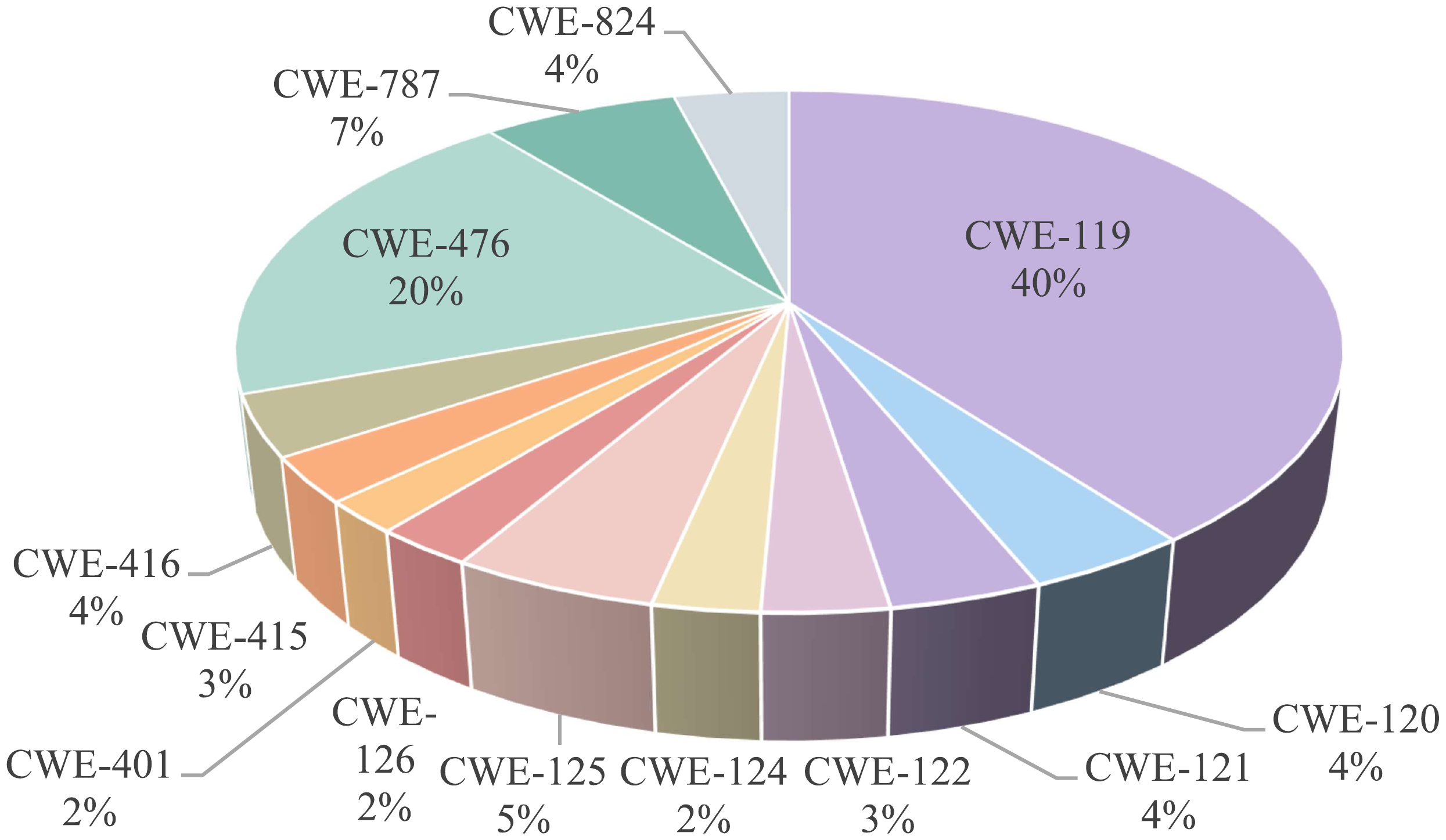}
\caption{Distribution of Vulnerability Types}
\label{Distribution}
\end{figure}

The statistics of the vulnerable programs in our dataset are shown in Table \ref{table1}. It includes 1,208 real-world vulnerabilities in CVE, covering 10 open-source C/C++ projects which are widely adopted as target projects by prior works \cite{Devign,VulDeePecker,SySeVR,DBLP:journals/corr/abs-2012-11701}, and 3,145 vulnerable samples (i.e., test cases) in SARD. Column 2 represents the scope of project versions affected by vulnerabilities in our dataset. Column 3 denotes the number of vulnerable samples. A vulnerable sample may contain one or more vulnerable functions. Column 4 and Column 5 are the number of nodes and edges in slices, respectively. Furthermore, the distribution of different types of memory-related vulnerabilities in our dataset is shown in Figure \ref{Distribution}, with CWE-119 (Improper Restriction of Operations within the Bounds of a Memory Buffer) accounting for the highest percentage at 40\% (including 1731 vulnerable samples). 

\subsubsection{Data Labeling.}
To train a detection model, we first need to conduct data labeling. There are two types of labels for statement nodes in the graph representation of a program: 1) \emph{vulnerable} represents that the node is related to an improper operation in the vulnerable programs; 2) \emph{non-vulnerable} represents that the node is related to the normal operation. To make this process automatic, we adopted a simple labeling strategy with \emph{diff} files. We first conduct program slicing for each vulnerable sample to generate a number of slices. Then, for each slice of the samples in SARD, we labeled the statement nodes annotated with "errors" as \emph{vulnerable}. For each slice of the real-world vulnerabilities in CVE, we compared statements in each slice and that in the corresponding vulnerable function according to \emph{diff} files. If a statement was deleted or altered (i.e., starting with "-" in \emph{diff} files), it would be labeled as \emph{vulnerable}, and \emph{non-vulnerable} otherwise. However, in practice,  part of memory-related vulnerabilities did not contain "-" in their patches. For example, in CVE-2019-19083 \cite{CVE-2019-19083}, memory leaks because allocated memory can not be released when memory allocation fails. This vulnerability can be fixed by adding a memory release statement. Thus, for these vulnerabilities can not be directly labeled, we manually labeled vulnerable nodes through identifying improper operations \cite{DBLP:conf/uss/LuPW19} (e.g., memory allocation or deallocation statements). In order to avoid introducing artificial deviation, two postgraduates and one Ph.D participated in this labeling process. If two postgraduates disagreed on the label of the same sample, the sample would be forwarded to the Ph.D evaluator for further investigation.

\subsubsection{Baseline Methods.} 
To answer the first research question, we selected three state-of-the-art DL-based vulnerability detection techniques, i.e., \emph{VulDeePecker} \cite{VulDeePecker}, \emph{SySeVR} \cite{SySeVR}, and \emph{Devign} \cite{Devign}. \emph{VulDeePecker} and \emph{SySeVR} represented source code as sequences and used BiLSTM model for vulnerability detection at the \emph{slice-level}. \emph{Devign} constructed a joint graph structure (including AST, CFG, DFG, and code sequences) and used GGNN model to detect vulnerabilities at the \emph{function-level}. They are widely adopted as baselines in recent works \cite{DeepWukong,FUNDED,IVDETECT} and have been shown to be effective in detecting memory-related vulnerabilities \cite{DeepWukong} even though they are designed for generic vulnerability detection.

To investigate RQ2, we selected  five  popular static analysis-based vulnerability detectors, i.e., \emph{PCA} \cite{PCA}, \emph{Saber} \cite{Saber1}, \emph{Flawfinder} \cite{Flawfinder}, \emph{RATS} \cite{RATS}, and \emph{Infer} \cite{Infer}. They have shown relatively good performance on memory-related vulnerabilities and are widely adopted as baselines in prior works \cite{DeepWukong,VulDeePecker,SySeVR,SMOKE}.

\subsubsection{Implementation}
We implemented \emph{MVD} in Python using PyTorch \cite{PyTorch}. Our experiments were performed  with the Nvidia Graphics Tesla T4 GPU, installed with Ubuntu 18.04, CUDA 10.1. 

The neural networks are trained in a batch-wise fashion until converging and the batch size is set to 32. The dimension of the vector representation of each node is set to 100 and the dropout is set to 0.1. ADAM \cite{ADAM} optimization algorithm is used to train the model with the learning rate of 0.001. Weight decay is set to 5$e$-1 and over-sampling scale is set as 1.0. The other hyper-parameters of our neural network are tuned through grid search. 

For RQ1, it is unfair to compare \emph{MVD} with other DL-based approaches because of our finer granularity. Thus, we used the \emph{function-level} as a compromise formula, i.e., if a vulnerable statement was identified correctly by \emph{MVD}, we would consider the function it belonged to was also detected correctly. For DL-based baselines, we respectively used the vulnerable and non-vulnerable functions as positive and negative samples to train the detection models. For \emph{MVD}, we only trained the detection model based on vulnerable functions (i.e., vulnerable statements are deemed as positive samples, while non-vulnerable statements are deemed as negative samples). We randomly chose 80\% of the programs for training and the remaining 20\% for testing. To make sure that our model was fine-tuned, we used \emph{ten-fold cross-validation} to evaluate the generalization ability of each approach. In RQ2, we evaluated \emph{MVD} and static analysis-based approaches at the \emph{function-level} and \emph{statement-level}, respectively. At the \emph{function-level}, we adopted the same experimental setup as RQ1. At the \emph{statement-level}, we evaluated each approach by randomly selecting 30 latest real-world vulnerabilities (reported in 2021) from our dataset, covering five common memory-related vulnerabilities (including \emph{Memory Leak} (ML), \emph{Double Free} (DF), \emph{Buffer Overflow} (BO), \emph{Use-After-Free} (UAF), and \emph{Out-of-bounds Read/Write} (OR/W)). These vulnerabilities are representative because they cover the vast majority of the vulnerability types in our dataset. For example, \emph{Buffer Overflow} (BO) corresponds to multiple CWEs, including CWE-119, -120, -121, and -122. To ensure that the trained model is tested on "unseen" programs, we excluded these samples from the training set of \emph{MVD}. For answering RQ3, we respectively replaced our FS-GNN model with three famous GNN models, including GCN \cite{GCN}, GGNN \cite{GGNN}, and RGCN \cite{RGCN}, to evaluate the contribution of each model to memory-related vulnerability detection. For answering RQ4, we recorded the average training and detection time of each approach in RQ1 and RQ2 to evaluate time cost of \emph{MVD} and baselines.

\subsection{Evaluation Metrics}
We used the following evaluation metrics to measure the effectiveness of our model:

\emph{\textbf{Accuracy (A)}} evaluates the performance that how many instances can be correctly labeled. It is calculated as:
\begin{equation}
    \boldsymbol{Accuracy} = \frac{TP+TN}{TP+FP+TN+FN}
\end{equation}

\emph{\textbf{Precision (P)}} is the fraction of true vulnerabilities among the detected ones. It is defined as:
\begin{equation}
    \boldsymbol{Precision} = \frac{TP}{TP+FP}
\end{equation}

\emph{\textbf{Recall (R)}} measures how many vulnerabilities can be correctly detected. It is calculated as:
\begin{equation}
    \boldsymbol{Recall} = \frac{TP}{TP+FN}
\end{equation}

\emph{\textbf{F1 score (F1)}} is the harmonic mean of $Recall$ and $Precision$, and can be calculated as:
\begin{equation}
    \boldsymbol{F1\ score} = 2\frac{Recall\cdot{Precision}}{Recall+Precision}
\end{equation}

\begin{table}[t]
 \caption{Comparison with DL-based approaches}
  \centering
  \begin{tabular}{ccccc}
    \toprule
    \textbf{Approach}     & \textbf{A ($\%$)}    & \textbf{P ($\%$)}    & \textbf{R ($\%$)}      & \textbf{F1 ($\%$)} \\
    \midrule
    \emph{VulDeePecker}   & 60.9  &  51.4    & 35.1     &  41.7  \\
    \emph{SySeVR}      & 63.4   & 53.3    & 62.9    &  57.7    \\
    \emph{Devign}      & 68.3    &  54.8   &  66.1     &  59.9    \\
    \emph{MVD}     & \textbf{74.1}    & \textbf{61.5}     & \textbf{69.4}    &  \textbf{65.2}     \\
    \bottomrule
  \end{tabular}
  \label{table2}
\end{table}

\section{Experimental Results}\label{Results}
\subsection{RQ1: \emph{MVD} VS. DL-Based Approaches}

Table \ref{table2} shows the overall results of each deep learning-based approach in terms of the aforementioned evaluation metrics. Overall, \emph{MVD} achieves better results and outperforms all of the three referred deep learning-based approaches. On average, for \emph{MVD}, the Accuracy is 74.1\%, the Precision is 61.5\%, the Recall is 69.4\%, and the F1 score is 65.2\%. Moreover, in terms of all metrics, \emph{MVD} can improve the best performed baseline \emph{Devign} by 5.0\%-12.2\%.

\begin{figure}
\centering
\begin{subfigure}{\linewidth}
\centering
\includegraphics[width=\linewidth]{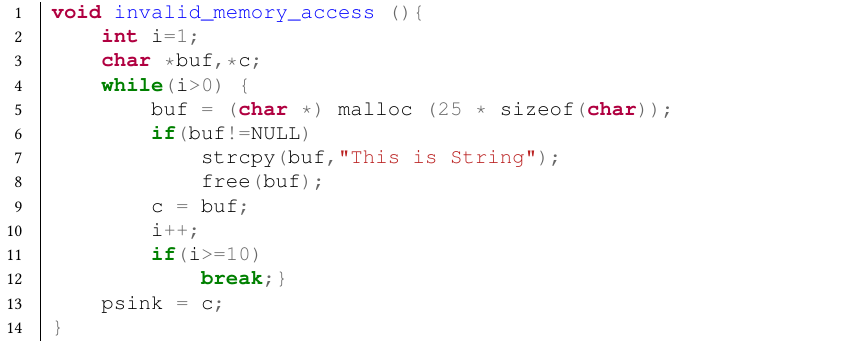}
\caption{A Vulnerability Missed by \emph{VulDeePecker}}\label{Sample1}
\quad
\end{subfigure}
\begin{subfigure}{\linewidth}
\centering
\includegraphics[width=\linewidth]{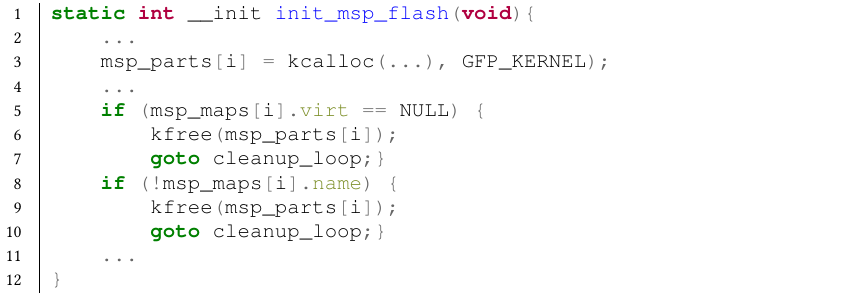}
\caption{A Non-vulnerable Code Sample Misidentified by \emph{SySeVR}}\label{Sample2}
\quad
\end{subfigure}
\begin{subfigure}{\linewidth}
\centering   
\includegraphics[width=\linewidth]{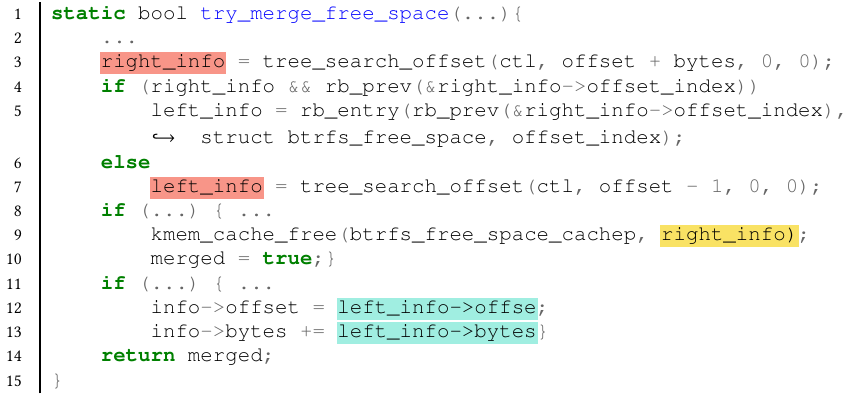}
\caption{CVE-2019-19448 \cite{CVE-2019-19448} Missed by \emph{Devign}}\label{Sample3}
\end{subfigure}
\caption{
\label{Sample} A Case study of RQ1}
\end{figure}

\underline{\emph{MVD} vs. \emph{VulDeePecker}.} As shown in Table \ref{table2}, our approach improves Accuracy, Precision, and F1 score over \emph{VulDeePecker} by 21.7\%, 19.6\%, and 56.4\%. Specifically, \emph{VulDeePecker} achieves the Recall of only 35.1\%. By contrast, the Recall of \emph{MVD} is as high as 69.4\%, nearly double (1.98x). Poor Recall indicates a great deal of vulnerabilities can not be detected by \emph{VulDeePecker}. A main reason is that \emph{VulDeePecker} only takes data-flow into account without regard to control-flow information. For example, as shown in Figure \ref{Sample1}, it contains a \emph{invalid memory access} vulnerability at \underline{line 9} because \emph{buf} has been freed at \underline{line 8} in an infinite \emph{while} loop. However, it is missed by \emph{VulDeePecker} in our experiment because without control-flow information, semantics of different branch structures will be ignored.

\underline{\emph{MVD} vs. \emph{SySeVR}.} We can observe from Table \ref{table2} that, in spite of significant improvement (1.79x) in Recall (62.9\%) in comparison with \emph{VulDeePecker}, \emph{SySeVR} still behaves worse than \emph{MVD} in terms of each metric. Particularly, \emph{MVD} improves Precision over \emph{SySeVR} by 15.4\%. The root cause for this performance gap is that \emph{SySeVR} cannot overcome the main limitation of sequence models (e.g., BiLSTM) in program semantics modeling. For example, in Figure \ref{Sample2}, there is a non-vulnerable code sample from \emph{Linux Kernel}, which is misidentified by \emph{SySeVR}. Although \emph{SySeVR} captures control- and data-dependence relations between statements by constructing Control Flow Graph (CFG) and Data Flow Graph (DFG), program semantics implied in these information can not be utilized because \emph{SySeVR} treats complex code structures as a sequential sequence of tokens, which omits the control flow divergence. Thus, this sample is misidentified as \emph{double free} because \texttt{msp\_parts[i]} is considered to be freed twice by \emph{kfree} at \underline{line 6} and \underline{line 9}.

\underline{\emph{MVD} vs. \emph{Devign}.} 
\emph{MVD} also outperforms the best performed baseline \emph{Devign}. It indicates that although powerful performance of GNN in inferring potential vulnerability semantics from graph representation of the program makes \emph{Devign} outstanding, the underutilization of flow information still restricts the performance of \emph{Devign} in detecting more complex memory-related vulnerabilities. For example, as shown in Figure \ref{Sample3}, it is a real-world vulnerability (CVE-2019-19448 \cite{CVE-2019-19448}) in \emph{Linux Kernel} missed by \emph{Devign}. It may lead to \emph{arbitrary address free} or \emph{double free} vulnerability by attacker because in certain situations, the pointer \texttt{left\_info} at \underline{line 7} can be the same as \texttt{right\_info} at \underline{line 3}. However, after \texttt{right\_info} has been freed at \underline{line 9}, \texttt{left\_info} which is the same as \texttt{right\_info} will be used at \underline{line 12-13} again. There are two main reasons why this vulnerability cannot be detected by \emph{Devign}. On the one hand, due to the lack of interprocedural analysis, precise program semantics like memory free (\texttt{kmem\_cache\_free} at \underline{line 9}) are hard to be captured by \emph{Devign}, causing imprecise semantic modeling. On the other hand, since \emph{Devign} processes multiple flows information by passing information in each individual graph and then aggregates them across graphs, \texttt{left\_info} and \texttt{right\_info} are treated as different data-flows, which causes complex semantic relations difficult to be preserved.


\begin{tcolorbox}
\textbf{Answer to RQ1:}\textit{ In comparison with the popular DL-based approaches, MVD achieves better detection performance by} fully utilizing flow information via interprocedural analysis and FS-GNN.

\end{tcolorbox}

\subsection{RQ2: \emph{MVD} VS. Static Analysis-Based Approaches} 

\begin{table}[t]
 \caption{Comparison with static analysis-based approaches}
  \centering
  \begin{tabular}{ccccc}
    \toprule
    \textbf{Approach}     & \textbf{A ($\%$)}    & \textbf{P ($\%$)}    & \textbf{R ($\%$)}      & \textbf{F1 ($\%$)} \\
    \midrule
    \emph{PCA}         & 65.2    & 48.9   & 61.1   & 54.3\\
    \emph{Saber}       & 64.4    & 47.6   & 59.2   & 52.8 \\
    \emph{Flawfinder}  & 61.1    & 18.2   & 23.5   & 20.5 \\
    \emph{RATS}        & 56.3    & 7.9    & 11.6   & 9.4  \\
    \emph{Infer}       & 50.7    & 33.1   & 54.8   & 41.3  \\
    \emph{MVD}          & \textbf{67.6}    & \textbf{54.8}     & \textbf{63.6}    &  \textbf{58.9} \\
    \bottomrule
  \end{tabular}
  \label{table3}
\end{table}

\begin{figure}
  \centering
  \includegraphics[width=\linewidth]{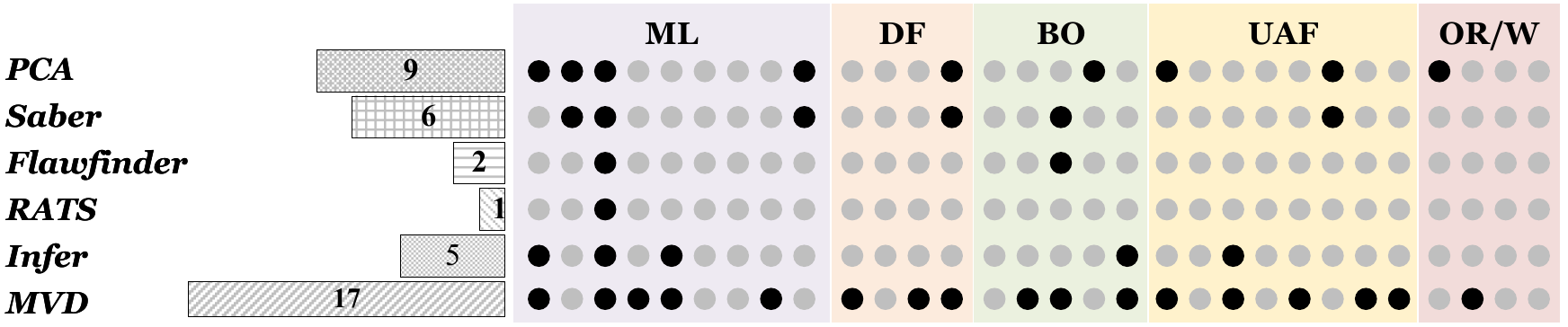}
\caption{The Number of Memory-Related Vulnerabilities in Real-World Projects Detected by Each Approach (ML: Memory Leak; DF: Double Free; BO: Buffer Overflow; UAF: Use-After-Free; OR/W: Out-of-bounds Read/Write)}
\label{5}
\end{figure}

Table \ref{table3} shows the experimental results of \emph{MVD} and the static analysis-based techniques. Overall, \emph{MVD} outperforms all baselines with regard to the evaluation metrics.

Among all static analysis-based baselines, \emph{PCA} and \emph{Saber} obtain relatively better detection performance.  \emph{PCA} achieves the highest Precision (48.9\%) and Recall (61.1\%) due to its consideration of global variables and accurate interprocedural flow analysis. 
Our approach still improves \emph{PCA} by 4.1\% in terms of Recall, and by 12.1\% in terms of Precision. The reason is that  static analysis-based approaches mainly rely on well-defined  vulnerability rules or patterns hand-crafted by human experts. They are effective in simple memory-related vulnerabilities (e.g., SARD dataset). However, real-world vulnerabilities are  more complicated, restricting the effectiveness of these static analysis-based detectors. Similar to these detectors, \emph{MVD} also  analyzes interprocedural control- and data-flow information. Owing to the powerful performance of deep learning models, \emph{MVD} can learn implicit vulnerability patterns from vulnerable code, instead of explicit rules or specifications, making it more effective in real-world scenarios.

\begin{figure}
  \centering
  \includegraphics[width=\linewidth]{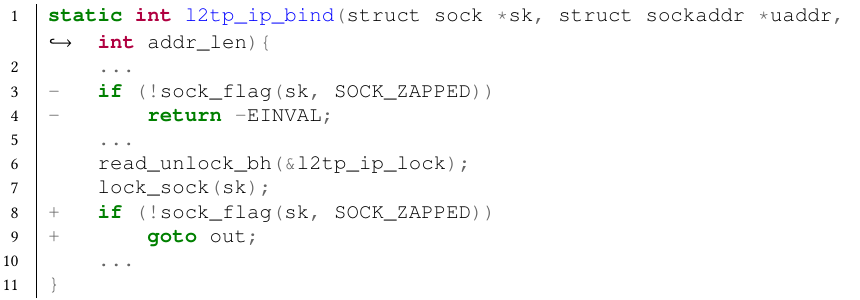}
\caption{A \emph{Use-After-Free} Vulnerability (CVE-2016-10200 \cite{CVE-2016-10200}) Missed by All Static Memory Detectors We Compared}
\label{sample4}
\end{figure}

Figure \ref{5} shows the detection results of each approach for five common memory-related vulnerabilities in real-world projects. These vulnerabilities are randomly selected from our dataset. Each detected individual vulnerability  successfully is labeled by a dark circle and the bars on the left-hand side are the total number of successfully detected vulnerabilities. Overall, \emph{MVD} outperforms all baselines by detecting 17 out of 30 vulnerabilities, including nine vulnerabilities cannot be detected by baselines.  Especially, in comparison with the best performed baseline \emph{PCA}, our approach can detect eight more vulnerabilities. For example, as shown in Figure \ref{sample4}, it is a \emph{use-after-free} vulnerability because a concurrent call could modify the socket flags between \texttt{sock\_flag(sk, SOCK\_ZAPPED)} at \underline{line 4} and \texttt{lock\_sock()} at \underline{line 8}, allowing local users to gain privileges or cause a denial of service by making multiple bind system calls without properly ascertaining whether a socket has the \texttt{SOCK\_ZAPPED} status. Unfortunately, it is missed by all the  static memory detectors because they cannot detect \emph{use-after-free} caused by race condition through static analysis only. In our approach, it can be correctly detected because of the advantage of deep learning models in mining implicit vulnerability patterns. 

\begin{tcolorbox}
\textbf{Answer to RQ2:}\textit{ With the advantage of deep learning models in mining implicit vulnerability patterns, MVD performs better in comparison with the popular static analysis-based approaches. }
\end{tcolorbox}

\subsection{RQ3: FS-GNN VS. Other GNNs} 
\begin{table}[t]
 \caption{ Contributions of different graph neural networks}
  \centering
  \begin{tabular}{ccccc}
    \toprule
    \textbf{Approach}     & \textbf{A ($\%$)}    & \textbf{P ($\%$)}    & \textbf{R ($\%$)}      & \textbf{F1 ($\%$)} \\
    \midrule
    \emph{GCN}       & 61.2   & 17.3  & 8.2   & 11.1 \\
    \emph{GGNN}      & 69.4   & 41.8  & 52.5  & 46.5\\
    \emph{RGCN}      & 72.7   & 49.3  & 58.1  & 53.3 \\
    \emph{FS-GNN}    & \textbf{77.5}     & \textbf{56.4}    & \textbf{62.9}    & \textbf{59.5}          \\
    \bottomrule
  \end{tabular}
  \label{table4}
\end{table}

Table \ref{table4} shows the results of different GNNs. We observe that FS-GNN can improve the best performed baseline RGCN by 6.7\%-14.4\%. There are mainly two reasons for this. First, FS-GNN adds edge types into the process of representation learning. It can be regarded as the joint learning of edge embedding and node embedding. Thus, FS-GNN can preserve the comprehensive program semantics based on interprocedural control- and data-flow, improving the flow-sensitivity for memory-related vulnerabilities. In addition, RGCN aggregates node and edge information through directed edge, while FS-GNN boosts the effect of edge types on context by adding corresponding inverse edges. Still taking the \emph{double free} vulnerability as an example, information of memory free in different branch statements will affect their condition nodes jointly. Therefore, important features of output nodes are also preserved by FS-GNN for node update and information propagation. In addition, we can find that although GGNN can process multiple relations across graphs, it is still limited by the increasing number of relations, resulting in  lower performance in comparison with RGCN and FS-GNN.

Furthermore, we observe that the performance of GCN is poor. The main reason is that the neglection of edge types leads to the missing of  structured code features (e.g., control- and data-flow). Without accurate control- and data-flow information, the performance of memory-related vulnerability detection drops sharply. 

\begin{table*}[h]
 \caption{Time cost in \underline{seconds} of different approaches. N/A: Not Applicable}
  \centering
  \begin{tabular}{ccccccccccc}
    \toprule
    \textbf{Method}     & \emph{MVD}    & \emph{VulDeePecker}    & \emph{SySeVR}      &\emph{Devign} &\emph{PCA}  &\emph{Saber} &\emph{Infer} &\emph{Flawfinder} &\emph{RATS} \\
    \midrule
    \textbf{Training Time(s)}   &  2386.2    &  1019.5    & 1833.9     & 2583.7 & N/A&N/A &N/A &N/A &N/A \\
    \textbf{Detection Time(s)}      & 10.4    &  8.1    & 9.7   &  11.9 & \textbf{9.2}   & 11.8 & 145.8  & 17.4   & 20.6 \\
    \bottomrule
  \end{tabular}
  \label{table5}
\end{table*}

\begin{tcolorbox}
\textbf{Answer to RQ3:}\textit{ FS-GNN can effectively contribute to the performance of MVD, as it can better capture the structured information of vulnerable code.}
\end{tcolorbox}

\subsection{RQ4: Efficiency}

Table \ref{table5} lists the time cost in seconds of each approach in training and detecting vulnerabilities. The results show that in comparison with the popular static analysis-based approaches, \emph{MVD} achieves less time cost over other approaches, except \emph{PCA}. This is because \emph{PCA} speeds up data dependence computation through sacrificing partial detection precision.

Among the four deep learning-based approaches, \emph{VulDeePecker} incurs the least training and detection time because it only considers data-flows and uses a simple sequence model, BiLSTM, for model training. However, combining with the results in Table \ref{table2}, we can find that it generates the lowest detection results because the lack of control-flows and the limitations of sequence model make it fail to capture the structured information.  Compared with other learning-based approaches, \emph{MVD} spends relatively longer training and detection time (excluding \emph{Devign}) because learning complicated program semantics in graphs is more time-consuming than in sequence. However, \emph{MVD} yields better detection results. In fact, due to the characteristic that deep learning models can be trained offline, their training cost may not be that important. Based on private vulnerability datasets, the users can train their own detection models offline and make a prediction within seconds.

\begin{tcolorbox}
\textbf{Answer to RQ4:}\textit{ In spite of a great deal of training time, MVD achieves relatively shorter detection time with better detection results, making a trade-off between accuracy and efficiency.} 

\end{tcolorbox}

\section{Threats to Validity}\label{Threats}
\paragraph{External validity.}
The main external threat to our study is the generalizability of our experiment results. We respectively investigated 4,353 vulnerable samples from 10 distinct C/C++ open-source projects and SARD, and used the mixed dataset for model evaluation like prior works. However, due to the huge gap in code complexity, detection results in practical scenarios may not be so satisfactory. Furthermore, our experiments are limited to memory-related vulnerabilities in C/C++ programs. Results may not be reproducible when applied to more complex vulnerabilities or languages (e.g., Java). Nevertheless, our approach is generic and can be extended for other vulnerabilities and languages.

\paragraph{Internal validity.}
Internal validity in our experiment relates to two factors. The first is our imperfect node labeling. In this work, we manually labeled nodes which did not contain any \emph{"delete"} statement as vulnerable through identifying related sensitive operations. Thus, it is possible that some samples are mislabeled. To avoid harmful influence caused by incorrect node labels, we tried our best to conduct the node labeling for the vulnerable samples in our dataset by three experienced researchers. In addition, the implementation of baselines also threats the results of our experiments. To compare with existing deep learning-based vulnerability detection approaches, we have re-implemented \emph{Devign} based on a popular repository\footnote{https://github.com/epicosy/devign} since it is closed-source. we try our best to build and tune the \emph{Devign} parameters on our dataset. 

\section{Related Works}\label{related work}
Existing memory-related vulnerability detection approaches can be divided into three main categories: static analysis-based, dynamic analysis-based, and learning-based approaches.

\textbf{Static Analysis-Based.} 
Static analysis-based approaches aim to detect vulnerabilities based on  specific vulnerability patterns or memory state model. Cherem et al. \cite{FastCheck} proposed a solution named \emph{FastCheck}, which reduces the memory leak analysis to a reachability problem over the guarded value-flow graph. Sui et al. \cite{Saber1,Saber2} proposed \emph{Saber}, a full-sparse value-flow graph (SVFG) based approach, to achieve the def-use chains and value-flow of the memory for pointer analysis. Shi et al. \cite{Pinpoint} proposed \emph{Pinpoint} to optimize widely-used sparse value-flow analysis through decomposing the cost of high-precision points-to analysis. Fan et al. \cite{SMOKE} presented \emph{SMOKE}, a staged approach for memory leak detection, to solve the scalability problem at industrial scale. 
Li et al. \cite{PCA} proposed \emph{PCA}, a static interprocedural data dependency analyzer, to speed up data dependency computation through partial call-path analysis. Differently, our approach can learn vulnerability features from large amounts of vulnerability data without requiring any prior knowledge of vulnerabilities.

\textbf{Dynamic Analysis-Based.}
Dynamic detection methods run the source code and dynamically track the allocation, use and release of memory at the run-time. \emph{LEAKPOINT} \cite{LEAKPOINT} monitored the state of memory objects based on stain analysis and tracked the last used location of memory and the location where references were lost. \emph{DoubleTake} \cite{DoubleTake} split the program execution into multiple blocks and saved the program state before each block started running. The program state would be checked after the execution of the block ended to judge whether there was an error in memory. \emph{Sniper} \cite{DBLP:conf/icse/JungLRP14} used the processor's monitoring unit (PMU) to track the access instructions to heap memory. Then, it calculated the staleness of heap objects and executed relevant instructions again to capture memory leakage during program execution. At the binary level, some dynamic analysis-based tools such as \emph{Valgrind} \cite{Valgrind}, \emph{Dr.Memory} \cite{Dr.Memory}, \emph{AddressSanitizer} \cite{ASAN} also perform well. They track memory allocation and deallocation during a program’s execution, and detect leaks by scanning the program’s heap for memory blocks that no pointer points to. Unlike dynamic analysis-based approaches, our approach does not require the execution of programs.

\textbf{Learning-Based.}
With the advance of machine learning (ML) and especially deep learning (DL) models, some approaches are proposed to automatically learn explicit or implicit vulnerability features from known vulnerabilities to identify unseen vulnerabilities in projects. Li et al. \cite{SySeVR,VulDeePecker} proposed two slice-based vulnerability detection approaches, \emph{VulDeePecker} and \emph{SySeVR}, to learn syntax and semantic information of vulnerable code. They represented source code as sequences at the \emph{slice-level} and used RNN (e.g., LSTM and BGRU) to train their detection models. Zou et al. \cite{mVulDeePecker} proposed an attention-based multi-class vulnerability detection approach, \emph{$\mu$VulDeePecker}, to pinpoint types of vulnerabilities. It introduced \emph{code attention} to accommodate information useful for learning local features and used a building-block BiLSTM to fuse different code features. Zhou et al. \cite{Devign} proposed a graph neural network-based vulnerability detection model through learning on a rich set of code semantic representations. 
Cheng et al. \cite{DeepWukong} embedded both textual and structured information of code into a comprehensive code representation and leveraged a GCN to perform the graph classification. Wang et at. \cite{FUNDED} proposed \emph{FUNDED}, a GNN-based vulnerability detection approaches. They combined nine mainstream graphs to extract finer program semantics and extended GGNN to model multiple code relationships. Different from existing learning-based vulnerability detection approaches, our approach aims to leverage rich flow information to support fine-grained memory-related vulnerability detection via the novel flow-sensitive graph neural networks.

\section{Conclusion}\label{Conclusion}

In this paper, we propose \emph{MVD} to detect  memory-related vulnerability  statements that are related to sensitive operations. \emph{MVD} employs a new graph neural network-based approach that leverages the flow-sensitive graph neural network (FS-GNN) to jointly embed both unstructured information and structured information for preserving high-level program semantics to learn implicit vulnerability patterns. The experimental results show the effectiveness of our approach  by comparing our approach with three state-of-the-art deep learning-based techniques and five popular static analysis-based memory detectors.

In the near future, we plan to compare our approach with more DL-based approaches (e.g., DeepWukong) and static memory detectors on a larger dataset to gain more insights. In addition, we aim  to investigate other code representation techniques to efficiently model flow information specific to memory-related vulnerabilities.

\begin{acks}
This work is supported by the National Natural Science Foundation of China (No.61872312, No.61972335, No.62002309, No.61902329);  the Six Talent Peaks Project in Jiangsu Province (No. RJFW-053), the Jiangsu ``333'' Project; the Natural Science Foundation of the Jiangsu Higher Education Institutions of China (No. 20KJB520016); the Open Funds of State Key Laboratory for Novel Software Technology of Nanjing University (No.KFKT2020B15, No.KFKT2020B16), the Yangzhou city-Yangzhou University Science and Technology Cooperation Fund Project (YZ2021157), and Yangzhou University Top-level Talents Support Program (2019).
\end{acks}
\balance
\bibliographystyle{ACM-Reference-Format}
\bibliography{sample-base}

\end{document}